\documentclass[10pt]{article}
\usepackage{amsmath}
\usepackage{amsfonts}

\usepackage{amsmath,amssymb}
\usepackage{graphicx}
\usepackage{color}
\usepackage{url}

\renewcommand{\arraystretch}{1.3}

\newtheorem{theorem}{Theorem}
\newtheorem{lemma}{Lemma}
\newtheorem{corollary}{Corollary}

\newtheorem{proposition}{Proposition}
\newtheorem{remark}{Remark}

\newtheorem{example}{Example}

\newcommand{\RNum}[1]{\uppercase\expandafter{\romannumeral #1\relax}}

\newcommand{\tr}{{{\rm Tr}}}

\newcommand{\ls}[1]
    {\dimen0=\fontdimen6\the\font\lineskip=#1\dimen0
     \advance\lineskip.5\fontdimen5\the\font
     \advance\lineskip-\dimen0
     \lineskiplimit=0.9\lineskip
     \baselineskip=\lineskip
     \advance\baselineskip\dimen0
     \normallineskip\lineskip\normallineskiplimit\lineskiplimit
     \normalbaselineskip\baselineskip
     \ignorespaces}

\begin{document}

\bibliographystyle{abbrv}

\title{Three-Weight Ternary Linear Codes from a Family of Monomials}
\author{Yongbo Xia \thanks{Y. Xia is with the Department of Mathematics and Statistics, South-Central University
			 for Nationalities, Wuhan 430074, China. Email: xia@mail.scuec.edu.cn},  
			Chunlei Li \thanks{C. Li is with the 
			 Department of Electrical Engineering and Computer Science, University of Stavanger, Stavanger, 4036, Norway. Email: Chunlei.Li@uis.no} 
	}

\date{}
\maketitle

\thispagestyle{plain} \setcounter{page}{1}

\begin{abstract}
Based on a generic construction, two classes of ternary  three-weight linear codes are obtained from a family of power functions, including some APN power functions. The weight distributions of these linear codes are determined through studying the properties of some exponential sum related to the proposed power functions.

\medskip


\smallskip

{\bf AMS } 94B15, 11T71
\end{abstract}

\ls{1.5}
\section{Introduction}\label{sec1}
%
%
%
%

Throughout this paper, we assume that $p$ is  an odd prime. For a
positive integer $m$, let $\mathbb{F}_{p^m}$ denote the finite field with
$p^m$ elements, $\mathbb{F}_{p^m}^*=\mathbb{F}_{p^m}\backslash\{0\}$ and $\alpha$
 a primitive element of $\mathbb{F}_{p^m}$.  Let $m$ and $k$ be two positive
integers such that ${m\over \gcd(k,m)}\geq 3$ is odd. Under these conditions, let $d$ be a positive integer
satisfying
\begin{equation}\label{cong}d\left(p^k+1\right)\equiv2\,\,\left({\rm mod}\,\,p^m-1\right).\end{equation}
Let \begin{equation}\label{dfset}D(a)=\left\{ x \in \mathbb{F}_{p^m}^*\,|\,{\rm Tr}_1^m(x^d)=a \right\},\,\,a\in \mathbb{F}_{p},\end{equation}
where  ${\rm Tr}^m_1(\cdot)$ is the trace function from $\mathbb{F}_{p^m}$ to
$\mathbb{F}_p$ \cite{LN}. Assume $D(a)$ contains $l_a$ different elements $\beta_1,\beta_2,\cdots,\beta_{l_a}$. For each $a\in \mathbb{F}_{p}$, we define a linear code of length $l_a$ over $\mathbb{F}_p$ by
\begin{equation}\label{dfcode}\mathcal{C}_{D(a)}=\left\{\left({\rm Tr}^m_1(\beta_1 x),{\rm Tr}^m_1(\beta_2 x),\cdots,{\rm Tr}^m_1(\beta_{l_a} x)\right): x \in \mathbb{F}_{p^m} \right\}\end{equation}
and call $D(a)$ the \textit{defining set} of this code.
In this paper, we study the linear codes $\mathcal{C}_{D(a)}$ defined by (\ref{cong})-(\ref{dfcode}) and prove the following two theorems.
\begin{theorem}\label{mainthem}
	For $p=3$, $\mathcal{C}_{D(0)}$ defined in (\ref{dfcode}) is a $[3^{m-1}-1,m]$ linear code with weight distribution given in Table \ref{tablein them 1}, where $e=\gcd(k,m)$.\end{theorem}

\begin{table*}[!t]
\renewcommand{\arraystretch}{1.3}
\caption{Weight distribution of $\mathcal{C}_{D(0)}$ in Theorem \ref{mainthem}} \label{tablein them 1}
\centering
\begin{tabular}{|c|c|}
\hline Weight  & Frequency \\
\hline $0$& $1$\\
\hline $2\cdot\left(3^{m-2}-3^{\frac{m+e}{2}-2}\right)$&
$3^{m-e}+3^{\frac{m-e}{2}}$\\
\hline$2\cdot\left(3^{m-2}+3^{\frac{m+e}{2}-2}\right)$&
$3^{m-e}-3^{\frac{m-e}{2}}$\\
\hline $2 \cdot3^{m-2}$& $3^m-1-2\cdot3^{m-e}$\\
\hline
\end{tabular}
\end{table*}

\begin{theorem}\label{mainthem1}

	For $p=3$ and $a\in\mathbb{F}_{3}^*$,
	(i)  if $e=\gcd(k,m)$ is even, or $e=\gcd(k,m)$ is odd and $d\equiv 1\,\,\left({\rm mod}\,\,3^e-1\right)$, then $\mathcal{C}_{D(a)}$ defined in (\ref{dfcode}) is a $[3^{m-1},m]$ linear code with weight distribution  given in Table \ref{tablein them 2};
	(ii) if $e=\gcd(k,m)$ is odd and $d\equiv 1+\frac{3^e-1}{2}\,\,\left({\rm mod}\,\,3^e-1\right)$, then $\mathcal{C}_{D(a)}$  is a  $[3^{m-1}+(-1)^{\frac{m-1}{2}}3^{\frac{m-1}{2}}\left(\frac{a}{3}\right),m]$ linear code and its possible nonzero weights are
\begin{equation}\label{weigt of D1} \arraycolsep=1.4pt\def\arraystretch{1.7}
\left \{\begin{array}{lll}
2\cdot\left(3^{m-2}+(-1)^{\frac{m-1}{2}}3^{\frac{m-3}{2}}\left(\frac{a}{3}\right)\right),\\
2\cdot\left(3^{m-2}+(-1)^{\frac{m-1}{2}}3^{\frac{m-3}{2}}\left(\frac{a}{3}\right)\right)\pm2\cdot3^{\frac{m-3}{2}},\\
2\cdot\left(3^{m-2}+(-1)^{\frac{m-1}{2}}3^{\frac{m-3}{2}}\left(\frac{a}{3}\right)\right)\pm\left(3^{\frac{m-3}{2}}-3^{\frac{m+e-4}{2}}\right),\\
2\cdot\left(3^{m-2}+(-1)^{\frac{m-1}{2}}3^{\frac{m-3}{2}}\left(\frac{a}{3}\right)\right)\pm\left(3^{\frac{m-3}{2}}+3^{\frac{m+e-4}{2}}\right),\\
2\cdot\left(3^{m-2}+(-1)^{\frac{m-1}{2}}3^{\frac{m-3}{2}}\left(\frac{a}{3}\right)\right)\pm\left(3^{\frac{m-3}{2}}+3^{\frac{m+2e-3}{2}}\right),
\end{array}\right.\end{equation}
where $\left(\frac{\cdot}{3}\right)$ denotes the quadratic character of $\mathbb{F}_3$.

\end{theorem}

\begin{table*}[!t]
\renewcommand{\arraystretch}{1.3}
\caption{Weight distribution of $\mathcal{C}_{D(a)}$  in Theorem \ref{mainthem1}} \label{tablein them 2}
\centering
\begin{tabular}{|c|c|}
\hline Weight  & Frequency \\
\hline $0$& $1$\\
\hline $2\cdot3^{m-2}-3^{\frac{m+e}{2}-2}$&
$3^{m-e}-3^{\frac{m-e}{2}}$\\
\hline$2\cdot3^{m-2}+3^{\frac{m+e}{2}-2}$&
$3^{m-e}+3^{\frac{m-e}{2}}$\\
\hline $2 \cdot3^{m-2}$& $3^m-1-2\cdot3^{m-e}$\\
\hline
\end{tabular}
\end{table*}

The idea of constructing linear codes from a defining set $D$ was introduced in \cite{Ding-luo-Niede2008, Ding-Niede2007}. Very recently several defining sets
	have been considered to generate
	linear codes with few weights \cite{Ding-Ding2015,tang2016 ie letter,qi2016 ie letter,tang2016it,zhou2016} . The defining sets therein are constructed from Bent functions and quadratic functions. The defining set in this paper is constructed from the general power functions with $d$ defined in (\ref{cong}), which covers several APN power functions (see Corollary \ref{coro2}).  The proposed linear codes also have applications in secrete sharing \cite{carlet-ding 2005,yuan-ding2006}, authentication codes \cite{ ding-wang2005}, association schemes \cite{Calderbank1984}, and strongly regular graphs \cite{Calderbank1984}.

The remainder of this paper is organized as follows. Section
\ref{pre} gives some preliminaries and notation, including some
useful lemmas. In Section \ref{sec3}, we calculate the weight distributions of two cyclic codes with three nonzero weights. Section \ref{con} concludes the study.

\section{Preliminaries}\label{pre}
Let $e$ be a divisor of a positive integer $m$. The trace
function from $\mathbb{F}_{p^m}$ to $\mathbb{F}_{p^e}$
is defined as
$$\tr^m_e(x)=\sum\limits_{i=0}^{\frac{m}{e}-1}x^{p^{ei}}.$$
It is well known that
$\tr^m_1(x)=\tr^e_1\left(\tr^m_e(x)\right)$ for any $x\in
\mathbb{F}_{p^m}$.

Throughout this paper we assume
that $q=p^e$ and $h=\frac{m}{e}$. Then $\mathbb{F}_{p^m}=\mathbb{F}_{q^h}$.  By identifying the finite field
$\mathbb{F}_{q^h}$ with the $h$-dimensional $\mathbb{F}_{q}$-vector
space $\mathbb{F}_{q}^h$, a function from $\mathbb{F}_{q^h}$ to
$\mathbb{F}_{q}$ can be regarded as an $h$-variable polynomial over
$\mathbb{F}_{q}$. In this sense, a function $f(x)$ from
$\mathbb{F}_{q^h}$ to $\mathbb{F}_{q}$ is called a {\it quadratic
form} over $\mathbb{F}_q$ if it can be written as a homogeneous polynomial in $\mathbb{F}_{q}[x_1,x_2,\cdots,x_h]$ of degree $2$ as
$$f\left(x_1,\cdots,x_h\right)=\sum\limits_{1\leq i\leq j\leq h}a_{ij}x_ix_j.$$
The rank $r$ of the quadratic form
$f(x)$ is defined as the codimension of the $\mathbb{F}_{q}$-vector
space
$$W=\left\{z\in \mathbb{F}_{q^h}\,|\,f(x+z)=f(x)\,\,{\rm for}\,\,{\rm
all}\,\,x\in \mathbb{F}_{q^h}\right\},$$ namely, $|W|=q^{h-r}$.

For each nonzero quadratic form $f(x)$ from $\mathbb{F}_{q^h}$ to
$\mathbb{F}_{q}$, there exists an $h\times h$ symmetric matrix $A$
such that $f(x)=X^{{\rm T}}AX,$
 where $X$ is written as a column vector and its transpose is $X^{{\rm
T}}=\left(x_{1}, x_{2},\cdots, x_{h}\right)\in \mathbb{F}_{q}^h$. By
Theorem 6.21 of \cite{LN}, there exists a nonsingular matrix $B$ of
order $h$ such that $B^{{\rm T}}AB$ is a diagonal matrix
$\mbox{diag}\left(a_1,a_2,\cdots,a_r,0,\cdots,0\right)$, where $r$
is the rank of the quadratic form $f(x)$ and $a_1,a_2,\cdots,a_r\in
\mathbb{F}_{q}^{*}$. By the
nonsingular linear substitution $X=BY$ with $Y^{{\rm
T}}=\left(y_{1}, y_{2},\cdots, y_{h}\right)$, the quadratic form $f(x)$ is transformed into  {\it a
diagonal form }  as
\begin{equation}\label{equivalent diag} f(x)=Y^{{\rm T}}B^{{\rm T}}ABY=
\sum\limits_{i=1}^{r}a_{i}y_i^2.
\end{equation}

 Given a positive integer $t$, let $\eta^{(t)}(\cdot)$ and $\chi^{(t)}(\cdot)$ denote
 the quadratic character and  the canonical
 additive character of $\mathbb{F}_{p^t}$, respectively. Namely,
\begin{equation}\label{multicha}
\eta^{(t)}(x)=\left \{
 \begin{array}{rll}
1, & \text{ if $x$ is a square in }\mathbb{F}_{p^t}^*,\\
-1, &\,\text{ if $x$ is a non-square in }\mathbb{F}_{p^t}^*,\\
0, &\,\text{ if } x=0,\\
\end{array} \right.
\end{equation} and
\begin{equation}\label{addcha}
\chi^{(t)}(x)=\omega_p^{{\rm Tr}^t_1(x)},\,\,x\in \mathbb{F}_{p^t},\end{equation}
where $\omega_p=e^{\frac{2\pi\sqrt{-1}}{p}}$ is a primitive complex $p$-th root of unity.
The following
two results related to Gaussian sums are useful in
the sequel.

\begin{lemma}\label{lemma guass}(\cite[Theorem 5.15 and Theorem 5.33 ]{LN} )
	Let $t$ be a positive integer and $\eta^{(t)}$ and $\chi^{(t)}$ defined in (\ref{multicha}) and (\ref{addcha}).
	 For any element $a$ in $\mathbb{F}_{p^t}^*$,
$$
\sum\limits_{x\in
\mathbb{F}_{p^t}}\omega_p^{\tr^t_1(ax^2)}=\eta(a)G(\eta^{(t)},\chi^{(t)}),
$$ where $G(\eta^{(t)},\chi^{(t)}) = \sum\limits_{x\in
	\mathbb{F}_{p^t}^*} \eta^{(t)}(x)\chi^{(t)}(x) $ is the  Gaussian sum  given by
\begin{equation}\label{GaussSum}
G(\eta^{(t)},\chi^{(t)})=\left \{
\begin{array}{lll}
(-1)^{t-1}p^{\frac{t}{2}}, &\,{\rm if}\,\,p\equiv1 \,\,({\rm mod}\,4),\\
(-1)^{t-1}(\sqrt{-1})^t p^{\frac{t}{2}},& \,{\rm if}\,\,p\equiv3 \,\,({\rm mod}\,4).\\
\end{array} \right.
\end{equation}
\end{lemma}

From Lemma \ref{lemma guass}, one can derive the following lemma immediately.

\begin{lemma}\label{lem1} (\cite[Lemma 1]{luo08}) Let $m=eh$ and $f(x)$ be a
quadratic form over $\mathbb{F}_{p^e}$ with rank
$r$ and the diagonal form  given in
(\ref{equivalent diag}).  Then,
\begin{align*}
&\sum\limits_{x\in \mathbb{F}_{p^m}}\omega_p^{\tr^e_1(f(x))}
=
\left \{
 \begin{array}{lll}
\eta^{(e)}(\Delta)(-1)^{(e-1)r}p^{m-\frac{er}{2}}, &\,{\rm for}\,\,p\equiv1 \,\,({\rm mod}\,4),\\
\eta^{(e)}(\Delta)(-1)^{(e-1)r}(\sqrt{-1})^{er}p^{m-\frac{er}{2}},& \,{\rm for}\,\,p\equiv3 \,\,({\rm mod}\,4),\\
\end{array} \right.
\end{align*}
where $\Delta=a_1a_2\cdots a_r$ with $a_1,a_2,\cdots,a_r$ in
(\ref{equivalent diag}).
\end{lemma}

  From Lemma \ref{lem1}, it follows that for any quadratic form $f(x)$ over $\mathbb{F}_q$
with rank $r$,
\begin{equation}\label{ppofquadritic}\sum\limits_{x\in
\mathbb{F}_{p^m}}\omega_p^{\tr^e_1\left(\lambda f(x)\right)}=\eta^{(e)}(\lambda^r)\sum\limits_{x\in
\mathbb{F}_{p^m}}\omega_p^{\tr^e_1\left(f(x)\right)},\,\,\forall \lambda
\in \mathbb{F}_{p^e}^*.\end{equation} Furthermore,
if $\lambda$ is a non-square in $\mathbb{F}_{p^e}$, one has
\begin{equation*}\label{Eq_SumOf2ExpSum}
	\begin{array}{rcl}
	\sum\limits_{x\in
		\mathbb{F}_{p^m}}\omega_p^{\tr_1^e\left(f(x)\right)}+\sum\limits_{x\in
		\mathbb{F}_{p^m}}\omega_p^{\tr_1^e\left(\lambda f(x)\right)}
	&=&
	\left \{
	\begin{array}{lll}
	\pm2p^{m-\frac{er}{2}},&\,\,r\,\,{\rm even},\\
	0,&\,\,r\,\,{\rm odd}.\\
	\end{array}
	\right.\\
	\end{array}
\end{equation*}

\vspace{1mm}

  Let $m$, $k$ be
two positive integers such that ${m \over e}$ is an odd integer
larger than $1$, where $e=\gcd(k,m)$.
Define\begin{equation}\label{Quv} Q_{u,v}(x)=\tr_e^m\left(u
x^{p^k+1}+v x^2\right),\quad u,\,v\in \mathbb{F}_{p^m},
\end{equation} and the associated exponential sum
\begin{equation}\label{Eq_T_uv}
T(u,v)=\sum\limits_{x\in \mathbb{F}_{p^m}
}\omega_p^{\tr^e_1\left(Q_{u,v}(x)\right)} = \sum\limits_{x\in \mathbb{F}_{p^m}
}\omega_p^{\tr^m_1\left(ux^{p^k+1}+vx^2\right)}.
\end{equation}
Note that when $u,v$ are not simultaneously zero,
$Q_{u,v}(x)$ is a nonzero quadratic form over $\mathbb{F}_{p^e}$.
The properties of $Q_{u,v}(x)$ and the associated exponential sum $T(u, v)$ have
been intensively studied in \cite{Muller99,H01,feng08,luo08,
xia10,tai-no2013,xia2015,licl2013, zhou-ding2014}. The following results are
useful in the sequel. \vspace{1mm}

\begin{lemma}\label{lem3}  Let $Q_{u,v}(x)$ be the
quadratic form defined by (\ref{Quv}),  where $(u,v)\in
\mathbb{F}_{p^m}^2\setminus \{(0,0)\}$ and  $h={m\over e}$.

(\textrm{i}) (\cite[Lemma 2]{luo08}) The rank of $Q_{u,v}(x)$ is
$h$, $h-1$ or $h-2$. Especially, both $Q_{u,0}(x)$ with $u\in
\mathbb{F}_{p^m}^*$ and $Q_{0,v}(x)$ with $v\in \mathbb{F}_{p^m}^*$
 have rank $h$.

(\textrm{ii}) (\cite[Lemma 6]{tai-no2013}
and \cite[Lemma 3.3]{zhou-ding2014} ) For any given $(u,v)\in
\mathbb{F}_{p^m}^2\setminus
 \{(0,0)\}$, at least one of $Q_{u,v}(x)$ and $Q_{u,-v}(x)$ has rank
 $h$.

\end{lemma}

\begin{lemma}\label{lem4}
(\cite[Theorem 1]{luo08}) Let $T(u,v)$ be the exponential sum defined in (\ref{Eq_T_uv}) and $\epsilon=\sqrt{\eta^{(e)}(-1)}$.  The value distribution of
$T(u,v)$ as
$(u,v)$ runs through $\mathbb{F}_{p^m}^2$
is given in Table \ref{table0}. Moreover, $T(u,v)=p^m$ if and only if $(u,v)=(0,0).$
\end{lemma}
\vspace{1mm}
\begin{table}[!t]\small
\renewcommand{\arraystretch}{1.3}
\caption{Value distribution for $T(u,v)$} \label{table0} \centering
\begin{tabular}{|c|c|}
\hline Value  & Frequency (each) \\
\hline $p^m$& $1$\\
\hline $\pm \epsilon
p^{\frac{m}{2}}$&$\frac{(p^m-1)p^{2e}(p^m-p^{m-e}-p^{m-2e}+1)}{2(p^{2e}-1)}$\\
\hline
$p^{\frac{m+e}{2}}$&$\frac{(p^m-1)(p^{m-e}+p^{\frac{m-e}{2}})}{2}$\\
\hline
$-p^{\frac{m+e}{2}}$&$\frac{(p^m-1)(p^{m-e}-p^{\frac{m-e}{2}})}{2}$\\
\hline $\pm \epsilon
p^{\frac{m+2e}{2}}$&$\frac{(p^m-1)(p^{m-e}-1)}{2(p^{2e}-1)}$\\
\hline
\end{tabular}
\end{table}

In addition, when $p^e\equiv
3\,\,({\rm mod}\,\,4)$, the value distribution of
$$\widehat{T}(u,v)=(T(u,v),T(-u,v))$$
as $(u,v)$ runs through $\mathbb{F}_{p^m}^*\times \mathbb{F}_{p^m}^*$ can be settled in the following lemma.
\begin{lemma}\label{lem6a}
For $p^e\equiv
3\,\,({\rm mod}\,\,4)$, the value distribution of $\widehat{T}(u,v)$ as $(u,v)$ runs through $\mathbb{F}_{p^m}^*\times \mathbb{F}_{p^m}^*$ is given in Table \ref{table in lem6a}, where $c_i$, $i=0,1,2$, are given by
\begin{equation}\label{c-value}c_i=\left \{
\begin{array}{lll}\sqrt{\eta^{(e)}(-1)}
p^{\frac{m+ie}{2}},&\,\,i=0,2,\\
p^{\frac{m+ie}{2}},&\,\,i=1.\\
\end{array}\right.
\end{equation}
\end{lemma}
{\em Proof:} When $p^e\equiv
3\,\,({\rm mod}\,\,4)$, $-1$ is a nonsquare in $\mathbb{F}_{p^e}^*$. Then, by Theorem 1 of \cite{xia2015}, the value distribution of $\widehat{T}(u,v)$ as $(u,v)$ runs through $\mathbb{F}_{p^m}\times \mathbb{F}_{p^m}$ can be obtained. Note that $\widehat{T}(0,0)=(p^m,p^m).$

Since $\frac{m}{\gcd(k,m)}$ is odd, $\gcd(p^k+1,p^m-1)=2$. Then, for any $u\in \mathbb{F}_{p^m}$, we have
\begin{equation*}\sum\limits_{x\in \mathbb{F}_{p^m}}\omega_p^{{\tr}^m_1\left(u x^{p^k+1}\right)}=\sum\limits_{x\in \mathbb{F}_{p^m}}\omega_p^{{\tr}^m_1\left(u x^2\right)}.\end{equation*} Combining this equality and Lemma \ref{lemma guass}, the value distribution of $\widehat{T}(u,0)$ as $u$ runs through $\mathbb{F}_{p^m}^*$ can be determined. Similarly, the value distribution of $\widehat{T}(0,v)$ as $v$ runs through $\mathbb{F}_{p^m}^*$ can also be derived.  Then, a straightforward calculation gives the desired result.
\hfill$\square$

\begin{table}[!t]\small
\renewcommand{\arraystretch}{1.3}
\caption{Value distribution of $\widehat{T}(u,v)$} \label{table in lem6a} \centering
\begin{tabular}{|c|c|}
\hline Value  & Frequency (each) \\
\hline
$\begin{array}{cc}(c_0,c_0)\\(-c_0,-c_0)\end{array}$&$\frac{(p^m-1)^2(p^e-3)}{4(p^{e}-1)}$\\

\hline $\begin{array}{cc}(-c_0,c_0),\,\,(c_0,-c_0)\end{array}$&$\frac{(p^{m}-1)[(p^{m}-1)(p^e-1)-4]}{4(p^e+1)}$\\

\hline $\begin{array}{cc}(c_0,c_1),\,\,(c_1,c_0)\\(-c_0,c_1),\,\,(c_1,-c_0)\end{array}$&$\frac{(p^m-1)(p^{m-e}+p^{\frac{m-e}{2}})}{4}$\\

\hline $\begin{array}{cc}(-c_0,-c_1),\,\,(-c_1,-c_0)\\(c_0,-c_1),\,\,(-c_1,c_0)\end{array}$&$\frac{(p^m-1)(p^{m-e}-p^{\frac{m-e}{2}})}{4}$\\

\hline $\begin{array}{cc}(c_0,c_2),\,\,(c_2,c_0)\\(-c_0,-c_2),\,\,(-c_2,-c_0)\end{array}$&$\frac{(p^m-1)(p^{m-e}-1)}{2(p^{2e}-1)}$\\

\hline $\begin{array}{cc}(-c_0,c_2),\,\,(c_2,-c_0)\\(c_0,-c_2),\,\,(-c_2,c_0)\end{array}$&$0$\\
\hline
\end{tabular}
\end{table}

\begin{lemma}\label{property of d}(\cite[Lemma 5]{xia2015}) Given $m$ and $k$ satisfying the condition that ${m\over \gcd(k,m)}\geq 3$ is odd, there are two distinct integers $d_1,d_2\in
	\mathbb{Z}_{p^m-1}$ satisfying (\ref{cong}), of which one satisfies $d\equiv
1\,\,({\rm mod}\,\,p^e-1)$, and the other
satisfies $d\equiv 1+\frac{p^e-1}{2}\,\,({\rm mod}\,\,p^e-1)$.
\end{lemma}

 In the sequel, we always assume that $\theta$ is a fixed non-square in $\mathbb{F}_{p^e}$. Then $\theta$
 is also a non-square in $\mathbb{F}_{p^m}$ since $\frac{m}{\gcd(m, k)}$ is odd. 
\begin{lemma}\label{lem nonsqure square} Denote by $\mathcal{S}$ the set of all square
 elements in $\mathbb{F}_{p^m}^*$. When $x$ runs
through $\mathbb{F}_{p^m}^*$ once, $x^{p^k+1}$ runs through
$\mathcal{S}$ twice. Moveover, $\mathbb{F}_{p^m}^*=\mathcal{S}\cup \theta\mathcal{S}$, where $\theta\mathcal{S}=\{\theta x\,|\,x\in
\mathcal{S}\}$.
\end{lemma}

\section{The weight distribution of $\mathcal{C}_{D(a)}$}\label{sec3}

In this section, we will give the proofs of Theorems \ref{mainthem} and \ref{mainthem1}.
Before we begin the proofs, we shall make some preparations.

Let $r_{u,v}$ denote the rank of the quadratic form $Q_{u,v}(x)$ defined in (\ref{Quv}). It follows from (\ref{ppofquadritic}) that for any  $\lambda\in \mathbb{F}_{p^e}^*$,
\begin{equation}\label{Eq_PropertyOfT_uv}
	T(\lambda u, \lambda v)
	= \eta^{(e)}(\lambda^{r_{u, v}})T(u, v)
	=
	\begin{cases}
	-T(u, v), &  \text{ if $\lambda$ is a non-square, $r_{u, v}$ is odd,} \\
	T(u, v), & \text{ otherwise.} 	\end{cases}
\end{equation}
This fact will be heavily used  in the sequel.

\begin{proposition}\label{prop0}

Let $T(u,v)$ be the exponential sum defined in  (\ref{Eq_T_uv}). Then, for each $\varepsilon\in \{1,-1\}$, the number of $u\in \mathbb{F}_{p^m}^*$ such that
$T(u,1)=\varepsilon p^{\frac{m+e}{2}}$ is equal to
$$\frac{p^{m-e}+\varepsilon p^{(m-e)/2}}{2}.$$
\end{proposition}
{\em Proof:}
By Lemmas \ref{lem1} and \ref{lem3}, if $T(u,v)=\varepsilon p^{\frac{m+e}{2}}$, the rank of ${\tr}_{e}^m(ux^{p^k+1}+vx^2)$ is $\frac{m}{e}-1$ and $(u,v)$ belongs to $\mathbb{F}_{p^m}^*\times \mathbb{F}_{p^m}^*$. For convenience,  with the notation introduced in Lemma \ref{lem nonsqure square}, we define the following notation:
\begin{equation*}\label{defsets}\begin{array}{lll}
M_{1,\varepsilon}=\left\{(u,v)\in \mathbb{F}_{p^m}^*\times \mathcal{S}:  T(u,v)=\varepsilon p^{\frac{m+e}{2}}\right\},\,\,M_{\theta,\varepsilon}=\left\{(u,v)\in \mathbb{F}_{p^m}^*\times \theta\mathcal{S}:  T(u,v)=\varepsilon p^{\frac{m+e}{2}}\right\},\\
N_{1,\varepsilon}=\left\{u\in \mathbb{F}_{p^m}^*:  T(u,1)=\varepsilon p^{\frac{m+e}{2}}\right\},\,\,N_{\theta,\varepsilon}=\left\{u\in \mathbb{F}_{p^m}^*:  T(u,\theta)=\varepsilon p^{\frac{m+e}{2}}\right\},\\\end{array}
\end{equation*}
where $\varepsilon\in \{1,-1\}$. Recall that $\mathbb{F}_{p^m}^*=\mathcal{S}\cup \theta\mathcal{S}$. By Lemma \ref{lem4}, one has
\begin{equation}\label{eq1-in pro0}
|M_{1,\varepsilon}|+|M_{\theta,\varepsilon}|=(p^m-1)\frac{p^{m-e}+\varepsilon p^{(m-e)/2}}{2},\,\,\varepsilon\in \{1,-1\}.
\end{equation}
In the following, we will prove two statements:

(\textrm{i}) $|N_{1,\varepsilon}|=|N_{\theta,\varepsilon}|$ for any $\varepsilon\in \{1,-1\}$;

(\textrm{ii}) $|M_{1,\varepsilon}|=\frac{p^m-1}{2}|N_{1,\varepsilon}|,\,\,|M_{\theta,\varepsilon}|=\frac{p^m-1}{2}|N_{\theta,\varepsilon}|$ for any $\varepsilon\in \{1,-1\}$.

For the first statement, let $u\in N_{1,\varepsilon}$, then we have $T(u,1)=\varepsilon p^{\frac{m+e}{2}}$ and the rank $r_{u,1}=\frac{m}{e}-1$ is even. It follows from (\ref{Eq_PropertyOfT_uv}) that
\begin{equation*}
\begin{array}{lll}T(\theta u,\theta)&=&
T(u,1).
\end{array}
\end{equation*}
Therefore, $u\in N_{1,\varepsilon}$ implies $u\theta\in N_{\theta,\varepsilon}$. Similarly, we can prove the converse: if $u\in N_{\theta,\varepsilon}$, then $\frac{u}{\theta}\in N_{1,\varepsilon}$. Thus, there is a one-to-one correspondence between these two sets and we have $|N_{1,\varepsilon}|=| N_{\theta,\varepsilon}|$ for each $\varepsilon\in \{1,-1\}$.

Now we prove the second statement.
Let $(u,v)\in M_{1,\varepsilon}$, then $v$ is a square element in $\mathbb{F}_{p^m}^*$ and we have
\begin{equation}\label{eq2-in pro0}T(u,v)=T\left(\frac{u}{v^{(p^k+1)/2}},1\right)\end{equation}
From (\ref{eq2-in pro0}), for each fixed $v\in \mathcal{S}$, the number of $u\in \mathbb{F}_{p^m}^*$ such that $T(u,v)=\varepsilon p^{\frac{m+e}{2}}$ is equal to
$|N_{1,\varepsilon}|$. Thus, we obtain $|M_{1,\epsilon}|=\frac{p^m-1}{2}|N_{1,\varepsilon}|$ for each $\varepsilon\in \{1,-1\}$. Similarly, one has $|M_{\theta,\epsilon}|=\frac{p^m-1}{2}|N_{\theta,\varepsilon}|$ for each $\varepsilon\in \{1,-1\}$.

The desired result follows from these two statements and (\ref{eq1-in pro0}). \hfill$\square$

\begin{proposition}\label{prop1}
Let $d$ be the integer satisfying (\ref{cong}) and
\begin{equation}\label{dfna}n_a=|\left\{x\in \mathbb{F}_{p^m}: {\tr}_1^m(x^d)=a\right\}|,\,\,a\in \mathbb{F}_{p}.\end{equation}
When $d$ satisfies $d\equiv
1\,\,\left({\rm mod}\,\,p^e-1\right)$,
$$n_a=p^{m-1};$$
When $d$ satisfies $d\equiv
1+\frac{p^e-1}{2}\,\,\left({\rm mod}\,\,p^e-1\right)$,
$$
n_a=\left \{
 \begin{array}{lll}
p^{m-1}, &\,{\rm if}\,\,p^e\equiv1 \,\,({\rm mod}\,4),\\
p^{m-1},&\,{\rm if}\,\,a=0\,\,{\rm and}\,\,p^e\equiv 3 \,\,({\rm mod}\,4),\\
p^{m-1}+(-1)^{\frac{m-1}{2}}p^{\frac{m-1}{2}}\eta^{(1)}(a),& \,{\rm if}\,\,a\neq 0\,\,{\rm and}\,\,p^e\equiv3 \,\,({\rm mod}\,4).\\
\end{array} \right.
$$
\end{proposition}

\noindent{\em Proof.} Using the theory of exponential sums, one can express $n_a$ as follows:
\begin{equation*}
\begin{array}{lll}n_a&=&\frac{1}{p}\sum\limits_{x\in \mathbb{F}_{p^m}}\sum\limits_{y\in \mathbb{F}_{p}}\omega_p^{y\left({\tr}_1^m(x^d)-a\right)}\\
&=&p^{m-1}+\frac{1}{p}\sum\limits_{y\in \mathbb{F}_{p}^*}\sum\limits_{x\in \mathbb{F}_{p^m}}\omega_p^{y\left({\tr}_1^m(x^d)-a\right)}\\
&=&p^{m-1}+\frac{1}{2p}\sum\limits_{y\in \mathbb{F}_{p}^*}\left[\sum\limits_{x\in \mathbb{F}_{p^m}}\omega_p^{y\left({\tr}_1^m(x^2)-a\right)}+\sum\limits_{x\in \mathbb{F}_{p^m}}\omega_p^{y\left({\tr}_1^m(\theta^d x^2)-a\right)}\right]\\
&=&p^{m-1}+\frac{1}{2p}\sum\limits_{y\in \mathbb{F}_{p}^*}\omega^{-ay}\left[\sum\limits_{x\in \mathbb{F}_{p^m}}\omega_p^{y{\tr}_1^m(x^2)}+\sum\limits_{x\in \mathbb{F}_{p^m}}\omega_p^{y{\tr}_1^m(\theta^d x^2)}\right]\\
&=&p^{m-1}+\frac{1}{2p}\sum\limits_{y\in \mathbb{F}_{p}^*}\omega^{-ay}\left[\eta^{(m)}(y)+\eta^{(m)}(\theta^d y)\right]G(\chi^{(m)},\eta^{(m)})\\
&=&p^{m-1}+\frac{1}{2p}G(\chi^{(m)},\eta^{(m)})\left[1+\eta^{(m)}(\theta^d )\right]\sum\limits_{y\in \mathbb{F}_{p}^*}\omega^{-ay}\eta^{(m)}(y)\\
&=&p^{m-1}+\frac{1}{2p}G(\chi^{(m)},\eta^{(m)})\left[1+\eta^{(m)}(\theta^d )\right]\sum\limits_{y\in \mathbb{F}_{p}^*}\omega^{-ay}\eta^{(e)}(y),\\
\end{array}
\end{equation*}
where the third and the fifth equalities hold due to Lemmas \ref{lem nonsqure square} and \ref{lemma guass}, respectively,  and the last equality holds since $\eta^{(m)}(x)=\eta^{(e)}(x)$ for any $x\in \mathbb{F}_{p^e}$. According to Lemma \ref{property of d}, we need to consider the following two cases.

\textit{Case 1:} $d\equiv
1\,\,\left({\rm mod}\,\,p^e-1\right)$. Then, $\theta^d=\theta$ and $\eta^{(m)}(\theta^d )=\eta^{(m)}(\theta)=-1$. Therefore, $n_a=p^{m-1}$ for any $a\in \mathbb{F}_p$.

\textit{Case 2:} $d\equiv
1+\frac{p^e-1}{2}\,\,\left({\rm mod}\,\,p^e-1\right)$. Then, $\theta^d=-\theta$ and one has
\begin{equation*}
\begin{array}{lll}n_a&=&p^{m-1}+\frac{1}{2p}G(\chi^{(m)},\eta^{(m)})\left[1+\eta^{(m)}(-\theta )\right]\sum\limits_{y\in \mathbb{F}_{p}^*}\omega^{-ay}\eta^{(e)}(y)\\
&=&p^{m-1}-\frac{1}{2p}G(\chi^{(m)},\eta^{(m)})\left[1-\eta^{(e)}(-1)\right]\sum\limits_{y\in \mathbb{F}_{p}^*}\omega^{-ay}\eta^{(e)}(y).
\end{array}
\end{equation*}
We consider the following two subcases.

\textit{Subcase 2.1:} $p^e\equiv
1\,\,{\rm mod}\,\,4$.  Then  $\eta^{(e)}(-1)=1$ and $n_a=p^{m-1}$ for any $a\in \mathbb{F}_p$.

\textit{Subcase 2.2:} $p^e\equiv
3\,\,{\rm mod}\,\,4$. Then $\eta^{(e)}(-1)=-1$ and $e$ must be odd. Thus, $\eta^{(e)}(y)=\eta^{(1)}(y)$ for any $y\in \mathbb{F}_{p}$ and
$$
\sum\limits_{y\in \mathbb{F}_{p}^*}\omega^{-ay}\eta^{(e)}(y)=\left \{
 \begin{array}{lll}
0, &\,{\rm if}\,\,a=0,\\
\eta^{(1)}(-a)G(\chi^{(1)},\eta^{(1)}), &\,{\rm if}\,\,a\neq 0.\\
\end{array} \right.
$$  Therefore, $n_a=p^{m-1}$ if $a=0$ and otherwise, $$n_a=p^{m-1}+\frac{1}{p}\eta^{(1)}(a)G(\chi^{(m)},\eta^{(m)})G(\chi^{(1)},\eta^{(1)}).$$
From (\ref{GaussSum}) in Lemma \ref{lemma guass}, the desired result follows.
 \hfill$\square$

 In the proof of Proposition \ref{prop1}, we actually have calculated the value distribution of the following exponential sums.
 \begin{corollary}\label{coro1}Let $ S(a)=\sum\limits_{y\in \mathbb{F}_{p}^*}\sum\limits_{x\in \mathbb{F}_{p^m}}\omega_p^{y\left({\tr}_1^m(x^d)-a\right)}$ and $n_a$ be defined in (\ref{dfna}). Then,
 $$S(a)=pn_a-p^{m}.$$
\end{corollary}

\begin{proposition}\label{prop2}
Let $T(u, v)$ be defined in (\ref{Eq_T_uv}) and $n_a$ given in Proposition \ref{prop1}. Define
\begin{equation}N(a,b)=|\left\{x\in \mathbb{F}_{p^m}: {\tr}_1^m(x^d)=a\,\,{\rm and}\,\,{\tr}_1^m(bx)=0\right\}|,\,\,a\in \mathbb{F}_{p},\,\, b\in \mathbb{F}_{p^m}^*.\end{equation}
If $d\equiv
1+\frac{p^e-1}{2}\,\,\left({\rm mod}\,\,p^e-1\right)$, $p^e\equiv
3\,\,{\rm mod}\,\,4$ and $a\in \mathbb{F}_{p}^*$, we have
\begin{equation}\label{eq0-res-pro3}
\begin{array}{lll}N(a,b)
=\frac{1}{p}n_a+\frac{1}{p^2}\sum\limits_{y\in \mathbb{F}_{p}^*}\omega_p^{-ya}
\sum\limits_{z\in \mathbb{F}_{p}^*}
T(yzb, y).
\end{array}
\end{equation}Otherwise,
\begin{equation}\label{eq-prop3}\begin{array}{lll}
N(a,b)=p^{m-2}+\frac{1}{2p^2}\left(\sum\limits_{z\in \mathbb{F}_{p}^*}
\big(T(zb, 1) + T(\theta zb, \theta)\big)\right)
\left(\sum\limits_{y\in \mathbb{F}_{p}^*}w_p^{-ay}\right),
\end{array}
\end{equation}  where $\theta$ is a non-square in $\mathbb{F}_{p^e}$.
\end{proposition}
{\em Proof:} Using the theory of exponential sums, $N(a,b)$ can be expressed as
\begin{equation}\label{eq0-prop3}
\begin{array}{lll}N(a,b)&=&\frac{1}{p^2}\sum\limits_{x\in \mathbb{F}_{p^m}}\sum\limits_{y\in \mathbb{F}_{p}}\omega_p^{y\left({\tr}_1^m(x^d)-a\right)}\sum\limits_{z\in \mathbb{F}_{p}}\omega_p^{z{\tr}_1^m(bx)}\\
&=&\frac{1}{p^2}\sum\limits_{y\in \mathbb{F}_{p}}\sum\limits_{z\in \mathbb{F}_{p}}\sum\limits_{x\in \mathbb{F}_{p^m}}\omega_p^{{\tr}_1^m\left(yx^d+zbx\right)-ya}\\
&=&p^{m-2}+\frac{1}{p^2}\sum\limits_{y\in \mathbb{F}_{p}^*}\sum\limits_{x\in \mathbb{F}_{p^m}}\omega_p^{{\tr}_1^m\left(yx^d\right)-ya}+\frac{1}{p^2}\sum\limits_{y\in \mathbb{F}_{p}^*}\sum\limits_{z\in \mathbb{F}_{p}^*}\sum\limits_{x\in \mathbb{F}_{p^m}}\omega_p^{{\tr}_1^m\left(yx^d+zbx\right)-ya}
\\&=& p^{m-2}+\frac{1}{p^2}S(a)+\frac{1}{p^2}R(a,b)
\\&=& \frac{1}{p}n_a+\frac{1}{p^2}R(a,b)
\end{array}
\end{equation}
where $S(a)$ is given by Corollary \ref{coro1} and
\begin{equation}\label{Eq_R_ab}
	R(a, b) = \sum\limits_{y\in \mathbb{F}_{p}^*}\omega_p^{-ya}\sum\limits_{z\in \mathbb{F}_{p}^*}\sum\limits_{x\in \mathbb{F}_{p^m}}\omega_p^{{\tr}_1^m\left(yx^d+zbx\right)}.
\end{equation}
By Lemma \ref{lem nonsqure square}, $R(a,b)$ can be represented as
\begin{equation}\label{Eq_Sum_Rab}
	R(a, b) = \frac{1}{2}\sum\limits_{y\in \mathbb{F}_{p}^*}\omega_p^{-ya}\sum\limits_{z\in \mathbb{F}_{p}^*}\sum\limits_{x\in \mathbb{F}_{p^m}}\left[\omega_p^{{\tr}_1^m\left(yx^2+zbx^{p^k+1}\right)}+\omega_p^{{\tr}_1^m\left(y\theta^d x^2+zb\theta x^{p^k+1}\right)}\right].
\end{equation}
According to Lemma \ref{property of d}, the exponential sum $R(a,b)$ will be investigated in the following two cases.

\textit{Case 1:} $\theta^d=\theta$. In this case, by Proposition \ref{prop1}, one has $n_a=p^{m-1}$. Moreover,
\begin{equation}\label{eq1-pro3}\def\arraystretch{2.2}
	\begin{array}{rcl}
	R(a,b)&=&\frac{1}{2}\sum\limits_{y\in \mathbb{F}_{p}^*}\omega_p^{-ya}\sum\limits_{z\in \mathbb{F}_{p}^*}\sum\limits_{x\in \mathbb{F}_{p^m}}\left[\omega_p^{{\tr}_1^m\left(yx^2+zbx^{p^k+1}\right)}+\omega_p^{{\tr}_1^m\left(y\theta x^2+zb\theta x^{p^k+1}\right)}\right]\\
	&=&\frac{1}{2}\sum\limits_{y\in \mathbb{F}_{p}^*}\omega_p^{-ya}\sum\limits_{u\in \mathbb{F}_{p}^*}\sum\limits_{x\in \mathbb{F}_{p^m}}\left[\omega_p^{{\tr}_1^m\left(uybx^{p^k+1}+yx^2\right)}+\omega_p^{{\tr}_1^m\left(uyb\theta x^{p^k+1}+y\theta x^2\right)}\right]
	\\
	&=&\frac{1}{2}\sum\limits_{y\in \mathbb{F}_{p}^*}\omega_p^{-ya}\sum\limits_{z\in \mathbb{F}_{p}^*}\sum\limits_{x\in \mathbb{F}_{p^m}}\left[\omega_p^{{\tr}_1^e\left(y{\tr}_e^m\left(zbx^{p^k+1}+x^2\right)\right)}+\omega_p^{{\tr}_1^e\left(y\theta{\tr}_e^m\left(zb x^{p^k+1}+x^2\right)\right)}\right]
	\\
	&=&\frac{1}{2}\sum\limits_{y\in \mathbb{F}_{p}^*}\omega_p^{-ya}\sum\limits_{z\in \mathbb{F}_{p}^*}
	\big(T(yzb, y)+T(y\theta zb, y\theta)\big).
	\end{array}
\end{equation}

For a given $y\in\mathbb{F}_{p}^*$, 
one of $y$ and $y\theta$ is a square in $\mathbb{F}_{p^e}^*$ and the other is a non-square since  $\eta^{(e)}(y)\eta^{(e)}(y\theta)=-1$. Thus, it follows from (\ref{Eq_PropertyOfT_uv})  that
\begin{equation}\label{eq2-pro3}
T(yzb, y)+T(y\theta zb, y\theta) = T(zb, 1)+T(\theta zb, \theta)
\end{equation}for given $b\in\mathbb{F}_{p^m}^*$, $y\in\mathbb{F}_{p}^*$ and $z\in\mathbb{F}_{p}^*$.
From (\ref{eq0-prop3}), (\ref{eq1-pro3}) and (\ref{eq2-pro3}), the desired result (\ref{eq-prop3}) in this case follows.

\textit{Case 2:} $\theta^d=-\theta$. In this case,
\begin{equation}\label{eq3-pro3}\def\arraystretch{2.2}
\begin{array}{rcl}R(a,b)
&=&\frac{1}{2}\sum\limits_{y\in \mathbb{F}_{p}^*}\omega_p^{-ya}\sum\limits_{z\in \mathbb{F}_{p}^*}\sum\limits_{x\in \mathbb{F}_{p^m}}\left[\omega_p^{{\tr}_1^m\left(yx^2+zbx^{p^k+1}\right)}+\omega_p^{{\tr}_1^m\left(-y\theta x^2+zb\theta x^{p^k+1}\right)}\right],\\
&=&\frac{1}{2}\sum\limits_{y\in \mathbb{F}_{p}^*}\omega_p^{-ya}\sum\limits_{u\in \mathbb{F}_{p}^*}\sum\limits_{x\in \mathbb{F}_{p^m}}\left[\omega_p^{{\tr}_1^m\left(yx^2+uybx^{p^k+1}\right)}+\omega_p^{{\tr}_1^m\left(-y\theta x^2-uyb\theta x^{p^k+1}\right)}\right]\\
&=&\frac{1}{2}\sum\limits_{y\in \mathbb{F}_{p}^*}\omega_p^{-ya}\sum\limits_{z\in \mathbb{F}_{p}^*}\sum\limits_{x\in \mathbb{F}_{p^m}}\left[\omega_p^{{\tr}_1^e\left(y{\tr}_e^m\left(zbx^{p^k+1}+x^2\right)\right)}+\omega_p^{{\tr}_1^e\left(-y\theta{\tr}_e^m\left(zb x^{p^k+1}+x^2\right)\right)}\right]
\\&=& \frac{1}{2}\sum\limits_{y\in \mathbb{F}_{p}^*}\omega_p^{-ya}\sum\limits_{z\in \mathbb{F}_{p}^*}
\big(T(yzb, y)+T(-y\theta zb, -y\theta)\big).\\
\end{array}
\end{equation}

\textit{Subcase 2.1:} $p^e\equiv
1\,\,{\rm mod}\,\,4$. Then, $-1$ is a square in $\mathbb{F}_{p^e}^*$.
It follows from Proposition \ref{prop1}  that $n_a=p^{m-1}$. Moreover, since $-\theta$ is a non-square in $\mathbb{F}_{p^e}^*$, we also have a result similar to (\ref{eq2-pro3}) as follows
$$T(yzb, y)+T(-y\theta zb, -y\theta) = T(zb, 1)+T(\theta zb, \theta)$$
for given $b\in\mathbb{F}_{p^m}^*$, $y\in\mathbb{F}_{p}^*$ and $z\in\mathbb{F}_{p}^*$.
This equality together with (\ref{eq3-pro3}) and (\ref{eq0-prop3}) leads to the desired result (\ref{eq-prop3}).

\textit{Subcase 2.2:} $p^e\equiv
3\,\,{\rm mod}\,\,4$ and $a=0$.  Then, $n_a=p^{m-1}$. By (\ref{eq3-pro3}), we have
\begin{equation*}\label{eq4-pro3}
\begin{array}{rcl}
R(a,b) &=&  \frac{1}{2}\sum\limits_{y\in \mathbb{F}_{p}^*}\sum\limits_{z\in \mathbb{F}_{p}^*}
\big(T(yzb, y)+T(-y\theta zb, -y\theta)\big)
\\&=&  \frac{1}{2}\sum\limits_{y\in \mathbb{F}_{p}^*}\sum\limits_{z\in \mathbb{F}_{p}^*}
T(yzb, y)+\frac{1}{2}\sum\limits_{u\in \mathbb{F}_{p}^*}\sum\limits_{z\in \mathbb{F}_{p}^*}T(u\theta zb, u\theta)
\\
&=& \frac{1}{2}\sum\limits_{y\in \mathbb{F}_{p}^*}\sum\limits_{z\in \mathbb{F}_{p}^*}
\big(T(yzb, y)+T(y\theta zb, y\theta)\big)
\\&= &\frac{1}{2}\left(\sum\limits_{z\in \mathbb{F}_{p}^*}
\big(T(zb, 1)+T(\theta zb, \theta)\big)\right) \left(\sum\limits_{y\in \mathbb{F}_{p}^*} 1\right),
\end{array}
\end{equation*}
where the last equality also follows from (\ref{eq2-pro3}). By (\ref{eq0-prop3}), the desired result (\ref{eq-prop3}) then follows.

\textit{Subcase 2.3:} $p^e\equiv
3\,\,{\rm mod}\,\,4$ and $a\in \mathbb{F}_{p}^*$. Then, $-1$ is a non-square in $\mathbb{F}_{p^e}$ and then $-\theta$ is a square element in $\mathbb{F}_{p^e}$.
By (\ref{Eq_PropertyOfT_uv}), we have $T(-y\theta zb, -y\theta ) = T(yzb, y)$. Thus,
\begin{equation*}\label{eq5-pro3}
\begin{array}{rcl}
R(a,b)&=&\frac{1}{2}\sum\limits_{y\in \mathbb{F}_{p}^*}\omega_p^{-ya}\sum\limits_{z\in \mathbb{F}_{p}^*}\big(
T(yzb, y) + T(yzb, y)\big)=\sum\limits_{y\in \mathbb{F}_{p}^*}\omega_p^{-ya}\sum\limits_{z\in \mathbb{F}_{p}^*}
T(yzb, y),
\end{array}
\end{equation*}This together with (\ref{eq0-prop3}) implies the desired result (\ref{eq0-res-pro3}).
\hfill$\square$

\medskip

With the above preparations, we can give the proofs of Theorems \ref{mainthem} and \ref{mainthem1}.

\noindent{\em Proof of Theorem \ref{mainthem}.} Let $\mathcal{C}_{D(0)}$ be the linear code defined in
(\ref{dfcode}) and
$$\mathbf{c}_b^0=\left({\rm Tr}^m_1(\beta_1 b),{\rm Tr}^m_1(\beta_2 b),\cdots,{\rm Tr}^m_1(\beta_{l_0} b)\right)\in \mathcal{C}_{D(0)}.$$
Note the length $l_0$ of this code is equal to $n_0-1$, where $n_0$ is given in Proposition \ref{prop1}. Denote the weight of $\mathbf{c}_b^0$ by $wt(\mathbf{c}_b^0)$.
It is obvious that $wt(\mathbf{c}_0^0)=0$. In the following, we assume that
$b\neq 0$. Then, using the notation in Propositions \ref{prop1} and \ref{prop2}, we have
\begin{equation}\label{eq1 in proof of them1}
\begin{array}{lll}wt(\mathbf{c}_b^0))&=&|\left\{x\in \mathbb{F}_{3^m}^*: {\tr}_1^m(x^d)=0 \right\}|-|\left\{x\in \mathbb{F}_{3^m}^*: {\tr}_1^m(x^d)=0\,\,{\rm and}\,\,{\tr}_1^m(bx)=0\right\}|\\
&=&|\left\{x\in \mathbb{F}_{3^m}: {\tr}_1^m(x^d)=0\right\}|-|\left\{x\in \mathbb{F}_{3^m}: {\tr}_1^m(x^d)=0\,\,{\rm and}\,\,{\tr}_1^m(bx)=0\right\}|\\
&=& n_0-N(0,b).
\end{array}
\end{equation}
When $p=3$, by Proposition \ref{prop2}, \begin{equation}\label{eq1-thm1}
\begin{array}{lll}N(0,b)&=&3^{m-2}+\frac{1}{2\cdot 3^2}\left(\sum\limits_{z\in \mathbb{F}_{3}^*}
\big( T(zb, 1) + T(\theta zb, \theta)\big)\right) \left(\sum\limits_{y\in \mathbb{F}_{3}^*} 1\right)\\
&=&3^{m-2}+\frac{1}{3^2} \Big[
T(b, 1)+ T(\theta b, \theta) + T(-b, 1) +T(-\theta b, \theta)
\Big].
\end{array}
\end{equation}
By (\ref{Eq_PropertyOfT_uv}), $T(b, 1)+ T(\theta b, \theta)\neq 0$  only if the rank
of the quadratic form $Q_{b, 1}(x)$ is even, i.e., $r_{b,1} = \frac{m}{e}-1$.
In this case,
$$
T(b, 1)+ T(\theta b, \theta) \in \left\{-2\cdot 3^{\frac{m+e}{2}}, 2\cdot 3^{\frac{m+e}{2}}\right\}.
$$ In a similar way, we have
$$
T(-b, 1)+ T(-\theta b, \theta) \in \left\{-2\cdot 3^{\frac{m+e}{2}}, 2\cdot 3^{\frac{m+e}{2}}\right\}
$$ only if the rank of $Q_{-b, 1}(x)$ equals $\frac{m}{e}-1$.

By Lemma \ref{lem3} (\textrm{ii}), at least one of the quadratic forms $Q_{b, 1}(x)$ and $Q_{-b, 1}(x)$ has rank $\frac{m}{e}$.
When one of $Q_{b, 1}(x)$ and $Q_{-b, 1}(x)$ has rank $\frac{m}{e}-1$, the other one must has rank $\frac{m}{e}$.
Consequently, the sums $T(b, 1)+ T(\theta b, \theta)$ and $T(-b, 1)+ T(-\theta b, \theta)$ cannot be
nonzero simultaneously.
Thus, by (\ref{eq1-thm1}), we have $$N(0,b)\in \left\{3^{m-2}-2\cdot 3^{\frac{m+e}{2}-2}, 3^{m-2}+2\cdot 3^{\frac{m+e}{2}-2}\right\}.$$

When $b$ runs through $\mathbb{F}_{3^m}^*$,
for each $\varepsilon \in\{1, -1\}$,
the number of $b$ such that $N(0,b)=3^{m-2}+2\varepsilon 3^{\frac{m+e}{2}-2}$ is equal to the number of $b$ such
that
$T(b, 1)=\varepsilon3^{\frac{m+e}{2}}$ or $T(-b, 1)=\varepsilon3^{\frac{m+e}{2}}$.  By Proposition \ref{prop0}, we can conclude the number of such $b$ is equal to $3^{m-e}+\varepsilon 3^{\frac{m-e}{2}}$, $\varepsilon \in \{1,-1\}$. This result together with (\ref{eq1 in proof of them1}) and Proposition \ref{prop1} gives the desired result.
\hfill$\square$

\medskip

\noindent{\em Proof of Theorem \ref{mainthem1}.}
For each $a\in \mathbb{F}_{3}^*$, let $\mathcal{C}_{D(a)}$ be the linear code defined in
(\ref{dfcode}) and $\mathbf{c}_b^a$ a codeword of $\mathcal{C}_{D(a)}$ given by
$$\left({\rm Tr}^m_1(\beta_1 b),{\rm Tr}^m_1(\beta_2 b),\cdots,{\rm Tr}^m_1(\beta_{n_a} b)\right).$$
Denote the weight of $\mathbf{c}_b^a$ by $wt(\mathbf{c}_b^a)$. It is easily seen that $wt(\mathbf{c}_0^a)=0$. In the sequel, we compute $wt(\mathbf{c}_b^a)$ for $b\neq 0$. By Proposition \ref{prop1}, the length of this code is $n_a$ and
\begin{equation}\label{eq1 in proof of them2}
\begin{array}{lll}wt(\mathbf{c}_b^a)&=&|\left\{x\in \mathbb{F}_{3^m}^*: {\tr}_1^m(x^d)=a \right\}|-|\left\{x\in \mathbb{F}_{3^m}^*: {\tr}_1^m(x^d)=a\,\,{\rm and}\,\,{\tr}_1^m(bx)=0\right\}|\\
&=&|\left\{x\in \mathbb{F}_{3^m}: {\tr}_1^m(x^d)=a\right\}|-|\left\{x\in \mathbb{F}_{3^m}: {\tr}_1^m(x^d)=a\,\,{\rm and}\,\,{\tr}_1^m(bx)=0\right\}|\\
&=& n_a-N(a,b).
\end{array}
\end{equation}
The following two cases are considered.

\textit{Case 1:} $e$ is even, or $e$ is odd and $d\equiv 1\,\,\left({\rm mod}\,\,3^e-1\right)$. Then by Propositions \ref{prop1} and \ref{prop2}, we have $n_a=3^{m-1}$ for each $a\in \mathbb{F}_3^*$ and
\begin{equation*}\begin{array}{lll}N(a,b)&=&3^{m-2}-\frac{1}{2\cdot3^2}\sum\limits_{z\in \mathbb{F}_{3}^*}
\big(T(zb, 1)+T(\theta zb, \theta) \big)\\
&=&3^{m-2}-\frac{1}{2\cdot 3^2}\Big[
\big(T(b, 1)+ T(\theta b, \theta)\big) + \big(T(-b, 1) +T(-\theta b, \theta)\big)
\Big].
\end{array}
\end{equation*}
A similar analysis as for (\ref{eq1-thm1}) in the proof of Theorem \ref{mainthem} yields the desired result.

\textit{Case 2:} $e$ is odd and $d\equiv 1+\frac{3^e-1}{2}\,\,\left({\rm mod}\,\,3^e-1\right)$. Note that $a\in \mathbb{F}_p^*$. Then, by  Proposition \ref{prop1}, we have $n_a=3^{m-1}+(-1)^{\frac{m-1}{2}}3^{\frac{m-1}{2}}\eta^{(1)}(a)$. By  Proposition \ref{prop2},
\begin{equation}\label{eq-six}\begin{array}{lll}N(a,b)
&=&\frac{1}{3}n_a+\frac{1}{3^2}\sum\limits_{y\in \mathbb{F}_{3}^*}\omega_3^{-ya}\sum\limits_{z\in \mathbb{F}_{3}^*}T(yzb, y)\\
&=&\frac{1}{3}n_a+\frac{1}{3^2}\,\omega_3^{-a}\big(T(b, 1)+ T(-b, 1)\big)+\frac{1}{3^2}\,\omega_3^{a}\big(T(-b, -1)+ T(b, -1)\big).
\end{array}
\end{equation} Note that $-1$ is a non-square in $\mathbb{F}_{3^e}^*$ in this case.
For each $b\in \mathbb{F}_{3^m}^*$,  if $\left(T(b,1),T(-b,1)\right)$ is given,
then the ranks of $Q_{b, 1}(x)$ and $Q_{-b, 1}(x)$ are determined. Consequently, by (\ref{Eq_PropertyOfT_uv}),
the value of  $\left(T(-b,-1),T(b,-1)\right)$ is uniquely determined. Then, by Lemma \ref{lem6a} and (\ref{eq-six}), we can calculate the possible values of $N(a,b)$ which are given in Table \ref{table in them2}, where $c_i$, $i=0,1,2$, are defined by (\ref{c-value}).
\begin{table}[!t]\small
\renewcommand{\arraystretch}{1.3}
\caption{Possible values  of $N(a,b)$} \label{table in them2} \centering
\begin{tabular}{|c|c|c|}
\hline $\left(T(b,1),T(-b,1)\right)$ & $\left(T(-b,-1),T(b,-1)\right)$& $N(a,b)$\\
\hline
$(c_0,c_0)$&$(-c_0,-c_0)$& $\frac{1}{3}n_a+\frac{2}{3^2}c_0(\omega_3^{-a}-\omega_3^{a})$\\
\hline
$(-c_0,-c_0)$&$(c_0,c_0)$& $\frac{1}{3}n_a+\frac{2}{3^2}c_0(\omega_3^{a}-\omega_3^{-a})$\\
\hline $\begin{array}{cc}(-c_0,c_0)\\(c_0,-c_0)\end{array}$&$\begin{array}{cc}(c_0,-c_0)\\(-c_0,c_0)\end{array}$&$\frac{1}{3}n_a$\\

\hline $\begin{array}{cc}(c_0,c_1)\\(c_1,c_0)\end{array}$&$\begin{array}{cc}(-c_0,c_1)\\(c_1,-c_0)\end{array}$
&$\frac{1}{3}n_a+\frac{c_0}{3^2}(\omega_3^{-a}-\omega_3^{a})+\frac{c_1}{3^2}(\omega_3^{-a}+\omega_3^{a})$\\

\hline $\begin{array}{cc}(-c_0,c_1)\\(c_1,-c_0)\end{array}$&$\begin{array}{cc}(c_0,c_1)\\(c_1,c_0)\end{array}$ &$\frac{1}{3}n_a+\frac{c_0}{3^2}(\omega_3^{a}-\omega_3^{-a})+\frac{c_1}{3^2}(\omega_3^{-a}+\omega_3^{a})$\\

\hline $\begin{array}{cc}(-c_0,-c_1)\\(-c_1,-c_0)\end{array}$&$\begin{array}{cc}(c_0,-c_1)\\(-c_1,c_0)\end{array}$
&$\frac{1}{3}n_a+\frac{c_0}{3^2}(\omega_3^{a}-\omega_3^{-a})-\frac{c_1}{3^2}(\omega_3^{-a}+\omega_3^{a})$\\

\hline $\begin{array}{cc}(c_0,-c_1)\\(-c_1,c_0)\end{array}$&$\begin{array}{cc}(-c_0,-c_1)\\(-c_1,-c_0)\end{array}$
&$\frac{1}{3}n_a+\frac{c_0}{3^2}(\omega_3^{-a}-\omega_3^{a})-\frac{c_1}{3^2}(\omega_3^{-a}+\omega_3^{a})$\\

\hline $\begin{array}{cc}(c_0,c_2)\\(c_2,c_0)\end{array}$&$\begin{array}{cc}(-c_0,-c_2)\\(-c_2,-c_0)\end{array}$
&$\frac{1}{3}n_a+\frac{c_0}{3^2}(\omega_3^{-a}-\omega_3^{a})+\frac{c_2}{3^2}(\omega_3^{-a}-\omega_3^{a})$\\

\hline $\begin{array}{cc}(-c_0,-c_2)\\(-c_2,-c_0)\end{array}$&$\begin{array}{cc}(c_0,c_2)\\(c_2,c_0)\end{array}$
&$\frac{1}{3}n_a+\frac{c_0}{3^2}(\omega_3^{a}-\omega_3^{-a})+\frac{c_2}{3^2}(\omega_3^{a}-\omega_3^{-a})$\\

\hline
\end{tabular}
\end{table}

Take $\omega_3=\frac{-1+\sqrt{-3}}{2}$. Then, by Table \ref{table in them2} and (\ref{eq1 in proof of them2}), the possible weights
of $\mathcal{C}_{D(1)}$ in (\ref{weigt of D1}) is obtained. Similarly, one can obtain the possible weights of $\mathcal{C}_{D(-1)}$ when $b\neq 0$.
\hfill$\square$
\begin{remark} When $p=3$ and $e=\gcd(k,m)$ is odd, the weight distribution of $\mathcal{C}_{D(a)}$ with $a\in \mathbb{F}_{p}^*$ is dependent on the value distribution of $\widehat{T}(b,1)=\left(T(-b,-1),T(b,-1)\right)$ as $b$ runs through $\mathbb{F}_{3^m}^*$. If we can determine the value distribution of $\widehat{T}(b,1)$, then the weight distribution of $\mathcal{C}_{D(a)}$ will be determined. Lemma \ref{lem6a} is necessary for determining the value distribution of $\widehat{T}(b,1)$. However, in order to find the value distribution of $\widehat{T}(b,1)$, we need to explore the relation between $\widehat{T}(b,1)$ and $\widehat{T}(u,v)$. It seems to be a difficult problem.
\end{remark}

As documented in  \cite{licl2013}, there are a number of integers satisfying
 the congruence (\ref{cong}), where three classes of APN exponents are also covered. From Theorem \ref{mainthem}, we have the following
corollary.
\begin{corollary}\label{coro2}Let $m\geq 3$ be odd and $x^d$ be an APN function over $\mathbb{F}_{3^m}$ with

\noindent \textrm{(i)} $d=\frac{3^m+1}{4}+\frac{3^m-1}{2}$ \cite{tor-anp-99}; or

\noindent \textrm{(ii)} $d=3^{\frac{m+1}{2}}-1$ \cite{tor-anp-99}; or

\noindent \textrm{(iii)} $d=\left(3^{\frac{m+1}{4}}-1\right)\left(3^{\frac{m+1}{2}}+1\right)$ for $m\equiv 3\,\, \mbox{mod}\,\,4$ \cite{cha-anp-2011}.\\
Then, these APN functions satisfy the congruence (\ref{cong}) and can be used for constructing $[3^{m-1}-1, m]$ linear codes $\mathcal{C}_{D(0)}$ with weight distribution given in Table \ref{tablein them 1}.

\end{corollary}

The following examples are provided for verifying the main results in Theorems \ref{mainthem} and \ref{mainthem1},  and they are confirmed by Magma.

\begin{example} Let $p=3$, $m=5$ and $k=2$.  Then, $e=\gcd(k,m)=1$, $p^e\equiv
3\,\,({\rm mod}\,\,4)$ and the congruence $d(p^k+1)\equiv 2\,\,({\rm
mod}\,\,p^m-1)$ has two solutions in $\mathbb{Z}_{p^m-1}$:
$d_1=97$, $d_2=218$. Using $d_1$ and $d_2$ in the construction given by (\ref{dfset}) and (\ref{dfcode}), the obtained linear code $\mathcal{C}_{D(0)}$ has length $80$ and the weight enumerator is $1+90x^{48}+80x^{54}+72x^{60}$.
\end{example}

\begin{example} Let $p=3$, $m=6$ and $k=2$.  Then, $e=\gcd(k,m)=2$, $p^e\equiv
1\,\,({\rm mod}\,\,4)$ and the congruence $d(p^k+1)\equiv 2\,\,({\rm
mod}\,\,p^m-1)$ has two solutions in $\mathbb{Z}_{p^m-1}$:
$d_1=73$, $d_2=437$. Using these two integers in the construction for $a=1$ or $a=2$, the obtained linear codes
 have length $243$ and they share the same weight enumerator $1+72x^{153}+566x^{162}+90x^{171}$.
\end{example}

\begin{example} Let $p=3$, $m=9$ and $k=3$.  Then, $e=\gcd(k,m)=3$, $p^e\equiv
3\,\,({\rm mod}\,\,4)$ and the congruence $d(p^k+1)\equiv 2\,\,({\rm
mod}\,\,p^m-1)$ has two solutions in $\mathbb{Z}_{p^m-1}$:
$d_1=703$, $d_2=10544$, where $d_1\equiv 1\,\,({\rm mod}\,\,p^e-1)$
and $d_2\equiv 1+\frac{p^e-1}{2}\,\,({\rm mod}\,\,p^e-1)$. Using $d_1$ in the construction for $a=1$ or $a=2$, the obtained linear codes have length $6561$ and they share the same weight enumerator $1+702x^{4293}+18224x^{4374}+756x^{4455}$.

Using $d_2$ in the construction for $a=1$ (resp. $a=2$), denote the obtained linear code by $\mathcal{C}_{1}$ (resp. $\mathcal{C}_{2}$). Then,  $\mathcal{C}_{1}$ has length $6642$ and its weight enumerator is
$$1+2x^{5184}+414x^{4536}+4848x^{4482}+9138x^{4428}+4938x^{4374}+342x^{4320},$$ and $\mathcal{C}_{2}$ has length $6480$ and the
weight enumerator is
$$1+2x^{3564}+342x^{4428}+4992x^{4374}+9138x^{4320}+4848x^{4266}+360x^{4212}.$$
 Note that not all the possible weights of $\mathcal{C}_{1}$ ( resp. $\mathcal{C}_{2}$) disappear.
\end{example}

\section{Conclusion}\label{con}

In this paper, two classes of three-weight ternary linear codes are obtained. Compared with the work in \cite{Ding-Ding2015,tang2016 ie letter,qi2016 ie letter,tang2016it,zhou2016}, we utilized
a family of power functions $x^d$, including three classes of APN power functions, to construct
the defining sets. We proceed our study mostly for general odd primes $p$, and obtained three-weight linear codes in the ternary case. The techniques and results in Propositions \ref{prop0}-\ref{prop2} would be useful for studying the weight distributions of other $p$-ary codes.

\section*{Acknowledgment}


\begin{thebibliography}{1}


\bibitem{Ding-luo-Niede2008} C. Ding, J. Luo, and H. Niederreiter, ``Two-weight codes punctured from
irreducible cyclic codes,"  in {\it Proc. 1st Int. Workshop Coding Theory
Cryptogr.}, 2008, pp. 119-124.
\bibitem{Ding-Niede2007} C. Ding and H. Niederreiter, ``Cyclotomic linear codes of order 3," {\it IEEE
Trans. Inform. Theory},  vol. 53, no. 6, pp. 2274-2277, Jun. 2007.

\bibitem{Ding-Ding2015} K. Ding, C. Ding, ``A class of two-weight and three-weight codes
and their applications in secret sharing",  {\it IEEE
Trans. Inform. Theory}, vol. 61, no. 11, pp. 5835-5842, Nov. 2015.

\bibitem{tang2016 ie letter}C. Tang, Y. Qi, D. Huang, ``Two-weight and three-weight linear codes from square functions", {\it IEEE Communications Letters}, vol. 20, no. 1, pp. 29-32, 2016.

\bibitem{qi2016 ie letter}Y. Qi, C. Tang, D. Huang, ``Binary linear codes with few weights", {\it IEEE Communications Letters}, vol. 20, no. 2, pp. 208-211, 2016.

\bibitem{tang2016it} C. Tang, N. Li, Y. Qi, Z. Zhou, T. Helleseth,``Linear codes with two or three weights from weakly regular Bent functions", {\it IEEE Trans. Information Theory }, vol. 62, no. 3, pp. 1166-1176, Mar. 2016.

\bibitem{zhou2016} Z. Zhou, N. Li, C. Fan, and T. Helleseth, ``Linear codes with two or
three weights from quadratic bent functions", {\it Designs, Codes Cryptogr.},
to be published. DOI 10.1007/s10623-015-0144-9.

\bibitem{Calderbank1984}  A. R. Calderbank and J. M. Goethals,``Three-weight codes and association
schemes", {\it Philips J. Res.}, vol. 39, nos. 4-5, pp. 143-152, 1984.


\bibitem{ding-wang2005}C. Ding, X. Wang,``A coding theory construction of new systematic
authentication codes", {\it Theoretical Comput. Sci.}, vol. 330, no. 1,
pp. 81-99, 2005.


\bibitem{carlet-ding 2005} C. Carlet, C. Ding, and J. Yuan, ``Linear codes from perfect nonlinear
mappings and their secret sharing schemes", {\it IEEE Trans. Inf. Theory},
vol. 51, no. 6, pp. 2089-2102, Jun. 2005.

\bibitem{yuan-ding2006}J. Yuan and C. Ding, ``Secret sharing schemes from three classes of
linear codes", {\it IEEE Trans. Inf. Theory}, vol. 52, no. 1, pp. 206-212,
Jan. 2006.

\bibitem{tor-gong2002}T. Helleseth and G. Gong, ``New nonbinary sequences with ideal
two-level autocorrelation," {\it IEEE Trans. Inf. Theory}, vol. 48,
no. 11,  pp. 2868-2872, Nov. 2002.

\bibitem{tang2005}X. H. Tang, P. Udaya, and P. Z. Fan, ``A new family of nonbinary
sequences with three-level correlation property and large linear
span," {\it IEEE Trans. Inf. Theory}, vol. 51, no. 8, pp. 2906-2914,
Aug. 2005.

\bibitem{xia2015}Y. Xia, C. Li, X. Zeng, T. Helleseth, ``Some results on cross-correlation distribution between a $p$-ary $m$-sequence and its decimated sequences," {\it IEEE Trans. Inf. Theory}, vol. 60, no. 11, pp. 7368-7381, Nov. 2014.


\bibitem{Muller99}E. N. M$\ddot{{\rm u}}$ller, ``On the crosscorrelation of sequences over $GF(p)$ with
short periods," {\it IEEE Trans. Inf. Theory}, vol. 45, no. 1, pp.
289-295, Jan. 1999.

\bibitem{LN}R. Lidl and H. Niederreiter,
``Finite Fields," in {\it Encyclopedia of Mathematics and Its
Applications}, vol. 20. Amsterdam, The Netherlands: Addison-Wesley,
1983.

\bibitem{xia10}Y. Xia, X. Zeng, and L. Hu, ``Further crosscorrelation properties of sequences
with the decimation factor $d=\frac{p^n+1}{p+1}-\frac{p^n-1}{2}$,"
{\it Appl. Algebra Eng. Commun. Comput.}, vol. 21, no. 5, pp.
329-342, 2010.
\bibitem{tai-no2013}S. T. Choi, J. Y. Kim, and J. S. No,``On the cross-correlation
of a $p$-ary $m$-sequence and its decimated sequences by
$d=\frac{p^n+1}{p^k+1}+\frac{p^n-1}{ 2}$," {\it IEICE Trans. Comm.},
vol. B96, no. 9, pp. 2190-2197, 2013.

\bibitem{fengluo07}K. Feng and J. Luo, ``Value distribution of exponential sums from
perfect nonlinear functions and their applications," {\it IEEE
Trans. Inf. Theory},  vol. 53, no. 9, pp. 3035-3041, Sept. 2007.

\bibitem{feng08}K. Feng and J. Luo, ``Weight distribution of some reducible cyclic
codes," {\it Finite Fields Appl.}, vol.14, no.4, pp. 390-409, Apr.
2008.

\bibitem{H01}Z. Hu, X. Li, D. Mills, E. N. M$\ddot{{\rm u}}$ller, W. Sun, W. Willems, Y. Yang and Z.
Zhang, ``On the crosscorrelation of sequences with the decimation
factor $d = \frac{p^n+1}{p+1} - \frac{p^n-1}{2}$," {\it Appl.
Algebra Eng. Commun. Comput.}, vol. 12, no. 3, pp. 255-263, 2001.

\bibitem{licl2013}C. Li, N. Li, T. Helleseth, and C. Ding,
``On the weight distributions of several classes of cyclic codes
from APN monomials,"  {\it IEEE Trans. Inf. Theory}, vol. 60, no. 8, pp. 4710-4721, Aug. 2014.

\bibitem{luo08}J. Luo and K. Feng, ``On the weight distribution of two classes of cyclic
codes," {\it IEEE Trans. Inf. Theory}, vol. 54, no. 12, pp.
5332-5344, Dec. 2008.
\bibitem{zhou-ding2014}Z. Zhou and C. Ding,``A class of three-weight cyclic codes," {\it Finite
Fields Appl.}, vol. 25, pp. 79-93, 2014.

\bibitem{tor-anp-99} T. Helleseth, C. Rong and D. Sandberg, ``New families of almost perfect nonlinear power mappings," {\it IEEE Trans. Inform.
Theory}, vol. 45, no. 2, pp. 475-485, Mar. 1999.


\bibitem{cha-anp-2011}Z. Zha and X. Wang, ``Almost perfect nonlinear power functions in odd characteristic," {\it IEEE Trans. Inform.
Theory}, vol. 57, no. 7, pp. 4826-4832, Jul. 2011.

































\end{thebibliography}
\end{document}